\documentclass[12pt,a4paper]{article}
\usepackage{times}
\usepackage{amsfonts,amsmath,amssymb}
\usepackage[a4paper]{geometry}
\usepackage{fancyhdr}
\usepackage{color}
\usepackage[pdftex]{hyperref}
\usepackage{graphicx}
\hypersetup{
    a4paper,
    breaklinks
}

\newtheorem{theorem}{Theorem}

\newtheorem{definition}[theorem]{Definition}
\newtheorem{example}[theorem]{Example}

\newcounter{ejmtFirstpage}
\usepackage[T1]{fontenc}
\usepackage{bbm}
\usepackage{bussproofs}
\usepackage{fancyvrb}
\usepackage{cprotect}
\usepackage{calc}

\def\atelierb{\textsf{Atelier~B}}

\def\bmth{\textsf{B}}
\def\cedric{\textsf{Cedric}}

\def\cnam{\textsf{Cnam}}
\def\coq{\textsf{Coq}}

\def\isabelle{\textsf{Isabelle}}
\def\isar{\textsf{Isar}}
\def\muscadet{\textsf{Muscadet}}
\def\siemens{\textsf{Siemens~IC-MOL}}

\def\szen{\textsf{Super~Zenon}}
\def\zenon{\textsf{Zenon}}

\def\brsep{\;|\;}
\def\bdots{\brsep{}\ldots{}\brsep{}}

\EnableBpAbbreviations 
\newcommand{\UICm}[1]{\UIC{$#1$}}
\newcommand{\AXCm}[1]{\AXC{$#1$}}
\newcommand{\BICm}[1]{\BIC{$#1$}}
\newcommand{\RLm}[1]{\RL{$#1$}}

\DefineVerbatimEnvironment{code}{Verbatim}{fontsize=\small}
\newcommand{\email}[1]{\href{mailto:#1}{\nolinkurl{#1}}}
\newlength{\codeindent}

\begin{document}

\title{Recovering Intuition from Automated Formal Proofs using Tableaux with
Superdeduction}

\author{\begin{tabular}{cc}
\textit{David~Delahaye} & \textit{Mélanie~Jacquel}\\
\email{David.Delahaye@cnam.fr} & \email{Melanie.Jacquel@siemens.com}\\
\cedric{}/\cnam{}, Paris, France & \siemens{}, Châtillon, France
\end{tabular}}

\date{}
\maketitle

\begin{abstract}
We propose an automated deduction method which allows us to produce proofs close
to the human intuition and practice. This method is based on tableaux, which
generate more natural proofs than similar methods relying on clausal forms, and
uses the principles of superdeduction, among which the theory is used to enrich
the deduction system with new deduction rules. We present two implementations of
this method, which consist of extensions of the \zenon{} automated theorem
prover. The first implementation is a version dedicated to the set theory of the
\bmth{} formal method, while the second implementation is a generic version able
to deal with any first order theory. We also provide several examples of
problems, which can be handled by these tools and which come from different
theories, such as the \bmth{} set theory or theories of the TPTP library.
\end{abstract}

\thispagestyle{fancy}


\section{Introduction}

These days, theorem proving appears as an appropriate support for education in
subjects such as mathematics and more generally logic, where proofs play a
significant role. This can be explained by the fact that some of the existing
theorem prover based systems have a long history of development, and constantly
provide technical innovations not only in terms of design, but also in terms of
theory. Among these theorem prover based systems, interactive theorem provers,
such as \coq{}~\cite{Coq} for example, appear to be quite appropriate tools,
since they offer special environments dedicated to proving. In particular, these
special environments offer syntax and type checking, as well as a bounded set of
tactics, i.e. commands building proofs when applied to proof goals. These
environments also provide some assistance in the way of building proofs, since
tactics are able to automatically and incrementally produce proofs when applied
to goals. This assistance does not only concern the application of tactics, but
may also be related to other aspects regarding modeling, such as the automated
generation of induction schemes from inductive types for instance. However,
these mechanized frameworks do not offer any guidance in the way of finding the
right proof of a theorem, and if the user does not have the intuition of this
proof (which may be acquired by thinking of the proof on paper), it is likely
that he/she would experience some difficulties in building the corresponding
proof, even with an interactive proof loop (even worse, he/she would probably
get lost by unnecessarily applying some tactics in an endless way, like
induction tactics for example). To deal with this problem of finding proofs, we
may consider the use of automated theorem provers as long as they at least
provide proof traces which are comprehensible enough to recover the intuition of
the corresponding proofs.

Automated theorem proving is a quite wide and still very active domain of
research. In automated theorem proving, we generally distinguish the semantic
methods from the syntactic methods. The semantic methods, such as the
Davis-Putnam algorithm~\cite{DP60} or the Binary Decision Diagrams~\cite{RB86}
(BDDs) for instance, have the advantage to be quite intuitive, but are limited
to propositional calculus. To deal with first order logic, we preferably rely on
syntactic methods, which may be split into two large families of methods. The
first family of methods is the saturation-based theorem proving, which was
actually introduced by Robinson with the resolution
calculus~\cite{JR65}. Resolution is a complete method working by refutation: a
contradiction (i.e. the empty clause) has to be deduced from an unsatisfiable
set of clauses. The search for a contradiction proceeds by saturating the given
set of clauses, that is, systematically (and exhaustively) applying all
applicable inference rules. The principle of resolution is general enough to
allow many calculi to be seen as resolution-based calculi (binary resolution,
positive resolution, semantic resolution, hyper-resolution, the inverse method,
etc). However, a proof produced by resolution is not appropriate to get the
intuition of the proof, since resolution actually works on a formula in clausal
form (a preliminary step therefore consists in clausifying the initial formula),
and there is little chance to understand the proof of the initial formula from
the resolution proof (unless the initial formula is already in clausal
form). The second family of syntactic methods tend to palliate this difficulty
and are called tableau-based methods. Tableaux are actually much older than
resolution-based methods and were introduced by pioneers Hintikka~\cite{KH55}
and Beth~\cite{EB55} from the cut-free version of Gentzen's sequent
calculus~\cite{GG35}. The tableau method still works by refutation but over the
initial formula contrary to resolution, and by case distinction. More precisely,
it allows us to systematically generate subcases until elementary contradictions
are reached, building a kind of tree from which it is possible to almost
directly produce a proof. Compared to resolution, tableaux therefore offer the
possibility to build comprehensible proofs which are directly related to the
corresponding initial formulas.

If tableaux allow us to produce more comprehensible proofs, some recent
deduction techniques have been developed and tend to improve the presentation of
proofs in the usual deductive systems, in particular when reasoning modulo a
theory. Among these new deduction techniques, there are, for example, deduction
modulo~\cite{DA03} and superdeduction~\cite{BA07}, which respectively focus on
the computational and deductive parts of a theory, and which can be considered
as steps toward high-level deductive languages. If deduction modulo and
superdeduction are equivalent when reasoning modulo a theory, superdeduction
appears to be more appropriate to produce proofs close to the human intuition as
it allows us to naturally encode custom deduction schemes. In addition, the
principle of superdeduction relies on the generation of specific deduction rules
(called superdeduction rules) from the axioms of the theory, and in practice, it
is quite easier to extend existing tools with ad hoc deduction rules than with a
congruence over the formulas (coming from the computational rules of deduction
modulo).

In this paper, we propose an automated deduction method based both on tableaux
and superdeduction. As said previously, the main motivation is to build a system
able to provide a significant help in matter of education by automatically
producing proofs comprehensible enough to recover the intuition of these
proofs. To show that such a system is actually effective in practice, we also
propose to implement this system by realizing an extension of an existing
automated theorem prover called \zenon{}~\cite{Zenon}, and which relies on
classical first order logic with equality and applies the tableau method as
proof search. In this context, the choice of \zenon{} is strongly influenced by
its ability of producing comprehensible proof traces (with several levels of
details). In addition, \zenon{} offers an extension mechanism, which allows us
to extend its core of deductive rules to match specific requirements, and which
is therefore quite appropriate to integrate superdeduction. Two extensions of
\zenon{} with superdeduction have been implemented and will be considered in
this paper. The first implementation is dedicated to the set theory of the
\bmth{} method~\cite{B-Book} (or \bmth{} for short), which is a formal method
and allows engineers to develop software with high guarantees of
confidence. This implementation is used by \siemens{} to automatically verify
\bmth{} proof rules coming from a database which is built adding rules from
their several projects and applications, such as driverless metro systems for
instance (see~\cite{JA11,JA12} for more details). The second implementation is
generic and works over any first order theory, which allows us to use it to
prove problems from the TPTP library~\cite{TPTP} (which is a library of test
problems for automated theorem proving systems).

The paper is organized as follows: in Section~\ref{sec:sded}, we first introduce
the principles of superdeduction; we then present, in Section~\ref{sec:auto},
the computation of superdeduction rules from axioms in the framework of the
tableau method used by \zenon{}; finally, in Sections~\ref{sec:bset}
and~\ref{sec:szen}, we respectively describe the implementation of our
extensions of \zenon{} for the \bmth{} set theory and for any first order
theory, and also provide some examples respectively coming from the database of
\bmth{} proof rules maintained by \siemens{} and the TPTP library.


\section{Principles of Superdeduction}
\label{sec:sded}

In this section, we present the principles of superdeduction, which is a variant
of deduction modulo, and which allows us to describe proofs in a more compact
format in particular. In addition, we show that proofs in superdeduction are not
only shorter, but also follow a more natural human reasoning scheme, and that
custom deduction schemes, such as structural induction over Peano natural
numbers for example, can be naturally encoded using superdeduction.

\subsection{Deduction Modulo and Superdeduction}

Deduction modulo~\cite{DA03} focuses on the computational part of a theory,
where axioms are transformed into rewrite rules, which induces a congruence over
propositions, and where reasoning is performed modulo this
congruence. Superdeduction~\cite{BA07} is a variant of deduction modulo, where
axioms are used to enrich the deduction system with new deduction rules, which
are called superdeduction rules. For instance, considering the inclusion in set
theory $\forall{}a,b\;(a\subseteq{}b\Leftrightarrow{}
\forall{}x\;(x\in{}a\Rightarrow{}x\in{}b))$, the proof of $A\subseteq{}A$ in
sequent calculus has the following form:

\begin{center}
\AXCm{}\RLm{\mathrm{Ax}}
\UICm{\ldots{},x\in{}A\vdash{}A\subseteq{}A,x\in{}A}
\RLm{{\Rightarrow}\mathrm{R}}
\UICm{\ldots{}\vdash{}A\subseteq{}A,x\in{}A\Rightarrow{}x\in{}A}
\RLm{\forall{}\mathrm{R}}
\UICm{\ldots{}\vdash{}A\subseteq{}A,\forall{}x\;(x\in{}A\Rightarrow{}x\in{}A)}
\AXCm{}\RLm{\mathrm{Ax}}
\UICm{\ldots{},A\subseteq{}A\vdash{}A\subseteq{}A}\RLm{{\Rightarrow}\mathrm{L}}
\BICm{\ldots{},
\forall{}x\;(x\in{}A\Rightarrow{}x\in{}A)\Rightarrow{}A\subseteq{}A\vdash{}
A\subseteq{}A}\RLm{\land{}\mathrm{L}}
\UICm{A\subseteq{}A\Leftrightarrow{}
\forall{}x\;(x\in{}A\Rightarrow{}x\in{}A)\vdash{}A\subseteq{}A}
\RLm{\forall{}\mathrm{L}\times{}2}
\UICm{\forall{}a,b\;(a\subseteq{}b\Leftrightarrow{}
\forall{}x\;(x\in{}a\Rightarrow{}x\in{}b))\vdash{}A\subseteq{}A}\DP
\end{center}

In deduction modulo, the axiom of inclusion can be seen as a computation rule
and therefore replaced by the rewrite rule
$a\subseteq{}b\rightarrow{}\forall{}x\;(x\in{}a\Rightarrow{}x\in{}b)$. The
previous proof is then transformed as follows:

\begin{center}
\AXCm{}\RLm{\mathrm{Ax}}
\UICm{x\in{}A\vdash{}x\in{}A}\RLm{{\Rightarrow}\mathrm{R}}
\UICm{\vdash{}x\in{}A\Rightarrow{}x\in{}A}
\RLm{\forall{}\mathrm{R}\mbox{, }
A\subseteq{}A\rightarrow{}\forall{}x\;(x\in{}A\Rightarrow{}x\in{}A)}
\UICm{\vdash{}A\subseteq{}A}\DP
\end{center}

It can be noticed that the proof is much simpler than the one completed using
sequent calculus. In addition to simplicity, deduction modulo also allows us for
unbounded proof size speed-up~\cite{GB11}.

Superdeduction proposes to go further than deduction modulo precisely when the
considered axiom defines a predicate $P$ with an equivalence
$\forall{}\bar{x}\;(P\Leftrightarrow{}\varphi{})$. While deduction modulo
replaces the axiom by a rewrite rule, superdeduction adds to this transformation
the decomposition of the connectives occurring in this definition. This
corresponds to an extension of Prawitz's folding and unfolding rules~\cite{DP65}
(called introduction and elimination rules by Prawitz), where the connectives of
the definition are introduced and eliminated. The proposed (right)
superdeduction rule is then the following (there is also a corresponding left
rule):

\begin{center}
\AXCm{\Gamma{},x\in{}a\vdash{}x\in{}b,\Delta{}}
\RLm{\mathrm{IncR}\mbox{, }x\not\in{}\Gamma{},\Delta{}}
\UICm{\Gamma{}\vdash{}a\subseteq{}b,\Delta{}}\DP
\end{center}

Hence, proving $A\subseteq{}A$ with this new rule can be performed as follows:

\begin{center}
\AXCm{}\RLm{\mathrm{Ax}}
\UICm{x\in{}A\vdash{}x\in{}A}\RLm{\mathrm{IncR}}
\UICm{\vdash{}A\subseteq{}A}\DP
\end{center}

This new proof is not only simpler and shorter than in deduction modulo, but
also follows a natural human reasoning scheme usually used in mathematics as
shown more precisely in the next subsection.

\subsection{Human Reasoning with Superdeduction}

Considering the previous example of inclusion in set theory, we can notice that
the superdeduction rule is more natural and intuitive than a simple folding rule
à la Prawitz. Given two sets $A$ and $B$, if we aim to prove $A\subseteq{}B$, it
seems a little unusual to propose to prove
$\forall{}x\;(x\in{}A\Rightarrow{}x\in{}B)$, instead we propose to prove
$x\in{}B$ given $x$ s.t. $x\in{}A$, which amounts to implicitly introducing the
connectives of the unfolded proposition. This implicit introduction of
connectives is precisely proposed by the previous superdeduction rule, which can
be read as ``if any element of $a$ is an element of $b$, then $a\subseteq{}b$''.

Similarly, superdeduction can also be used to naturally encode custom deduction
schemes. For example, let us consider the structural induction scheme over Peano
natural numbers (i.e. non-negative integers). This scheme can be defined as
follows (i.e. the natural numbers are seen as the set of terms verifying the
inductive predicate):

$$\forall{}n\;(n\in{}\mathbbm{N}\Leftrightarrow{}\forall{}P\;
(0\in{}P\Rightarrow{}\forall{}m\;(m\in{}P\Rightarrow{}S(m)\in{}P)
\Rightarrow{}n\in{}P))$$

In sequent calculus, this scheme can be encoded by the two following (right)
superdeduction rules (there are also two corresponding left rules):

\begin{center}
\AXCm{\Gamma{},0\in{}P,H(P)\vdash{}n\in{}P,\Delta{}}
\RLm{\mathrm{IndR},P\not\in{}\Gamma{},\Delta{}}
\UICm{\Gamma{}\vdash{}n\in{}\mathbbm{N},\Delta{}}\DP
\end{center}

\begin{center}
\AXCm{\Gamma{},m\in{}P\vdash{}S(m)\in{}P,\Delta{}}
\RLm{\mathrm{HeredR},m\not\in{}\Gamma{},\Delta{}}
\UICm{\Gamma{}\vdash{}H(P),\Delta{}}\DP
\end{center}

Let us notice that the induction scheme actually requires two superdeduction
rules, whose one of the rules (the rule $\mathrm{HeredR}$) focuses on the
heredity part of the scheme in particular. This focus is motivated by the need
of avoiding permutability problems (between Skolemization and instantiation),
which may occur when computing superdeduction rules. These permutability
problems are quite common in automated proof search, and appear here since
superdeduction systems are in fact embedding a part of compiled automated
deduction. In~\cite{BA07} and in order to deal with these permutability
problems, the authors use a method inspired by focusing techniques in the
framework of sequent calculus. It is worth noting that these permutability
problems are managed in different ways by automated deduction methods, and in
particular, we will therefore not have to use focusing techniques when
integrating superdeduction to the tableau method in Section~\ref{sec:auto}.


\section{Tableaux with Superdeduction}
\label{sec:auto}

In this section, we present the tableau method used by the \zenon{} automated
theorem prover, which deals with classical first order logic with equality and a
specific support for equivalence relations. Once the rules of this method have
been described, we show how it is possible to compute superdeduction rules from
axiomatic theories, and how these new rules extend the kernel of rules of
\zenon{}.

\subsection{The Tableau Method}

The proof search rules of \zenon{} are described in detail in~\cite{Zenon} and
summarized in Figure~\ref{fig:zenon} (for the sake of simplification, the
unfolding and extension rules are omitted), where the ``$|$'' symbol is used to
separate the formulas of two distinct nodes to be created, $\epsilon$ is
Hilbert's operator ($\epsilon(x).P(x)$ means some $x$ that satisfies $P(x)$, and
is considered as a term), capital letters are used for metavariables, and $R_r$,
$R_s$, $R_t$, and $R_{ts}$ are respectively reflexive, symmetric, transitive,
and transitive-symmetric relations (the corresponding rules also apply to the
equality in particular). As hinted by the use of Hilbert's operator, the
$\delta{}$-rules are handled by means of $\epsilon{}$-terms rather than using
Skolemization. What we call here metavariables are often named free variables in
the tableau-related literature; they are not used as variables as they are never
substituted. The proof search rules are applied with the normal tableau method:
starting from the negation of the goal, apply the rules in a top-down fashion to
build a tree. When all branches are closed (i.e. end with an application of a
closure rule), the tree is closed, and this closed tree is a proof of the
goal. This algorithm is applied in strict depth-first order: we close the
current branch before starting work on another branch. Moreover, we work in a
non-destructive way: working on one branch will never change the formulas of
another branch.

\begin{figure}[htbp]
\framebox[\textwidth][c]
{\parbox{\textwidth}
{\small
\hspace{0.2cm}\underline{Closure and Cut Rules}
\begin{center}
\begin{tabular}{c@{\hspace{1cm}}c@{\hspace{1cm}}c}    
\AXCm{\bot}\RLm{\odot_\bot}\UICm{\odot}\DP &
\AXCm{\neg\top}\RLm{\odot_{\neg\top}}\UICm{\odot}\DP &
\AXCm{}\RL{cut}\UICm{P\brsep \neg P}\DP\\\\
\AXCm{\neg{}R_r(t,t)}\RLm{\odot_r}\UICm{\odot}\DP &
\AXCm{P}\AXCm{\neg{}P}\RLm{\odot}\BICm{\odot}\DP &
\AXCm{R_s(a,b)}\AXCm{\neg{}R_s(b,a)}\RLm{\odot_s}
\BICm{\odot}\DP
\end{tabular}
\end{center}
\hspace{0.2cm}\underline{Analytic Rules}
\begin{center}
\begin{tabular}{c@{\hspace{1cm}}c@{\hspace{1cm}}c}
\AXCm{\neg\neg{}P}\RLm{\alpha_{\neg\neg}}
\UICm{P}\DP &
\AXCm{P\Leftrightarrow{}Q}\RLm{\beta_\Leftrightarrow}
\UICm{\neg{}P,\neg{}Q\brsep{}P,Q}\DP &
\AXCm{\neg(P\Leftrightarrow{}Q)}\RLm{\beta{\neg\Leftrightarrow}}
\UICm{\neg{}P,Q\brsep{}P,\neg{}Q}\DP\\\\
\AXCm{P\land{}Q}\RLm{\alpha_\land}\UICm{P,Q}\DP &
\AXCm{\neg(P\vee{}Q)}\RLm{\alpha_{\neg\vee}}
\UICm{\neg{}P,\neg{}Q}\DP &
\AXCm{\neg(P\Rightarrow{}Q)}\RLm{\alpha_{\neg\Rightarrow}}
\UICm{P,\neg{}Q}\DP\\\\
\AXCm{P\vee{}Q}\RLm{\beta_\vee}
\UICm{P\brsep{}Q}\DP &
\AXCm{\neg(P\land{}Q)}\RLm{\beta_{\neg\land}}
\UICm{\neg{}P\brsep\neg{}Q}\DP &
\AXCm{P\Rightarrow{}Q}\RLm{\beta_\Rightarrow}
\UICm{\neg{}P\brsep{}Q}\DP\\\\
\multicolumn{3}{c}{
\begin{tabular}{c@{\hspace{1cm}}c}
\AXCm{\exists{}x\;P(x)}\RLm{\delta_\exists}
\UICm{P(\epsilon(x).P(x))}\DP &
\AXCm{\neg\forall{}x\;P(x)}\RLm{\delta_{\neg\forall}}
\UICm{\neg{}P(\epsilon(x).\neg{}P(x))}\DP
\end{tabular}}
\end{tabular}
\end{center}
\hspace{0.2cm}\underline{$\gamma$-Rules}
\begin{center}
\begin{tabular}{c@{\hspace{1cm}}c@{\hspace{1cm}}c@{\hspace{1cm}}c}
\AXCm{\forall{}x\;P(x)}\RLm{\gamma_{\forall{}M}}
\UICm{P(X)}\DP &
\AXCm{\neg\exists{}x\;P(x)}\RLm{\gamma_{\neg\exists{}M}}
\UICm{\neg{}P(X)}\DP &
\AXCm{\forall{}x\;P(x)}\RLm{\gamma_{\forall\mathrm{inst}}}
\UICm{P(t)}\DP &
\AXCm{\neg\exists{}x\;P(x)}\RLm{\gamma_{\neg\exists\mathrm{inst}}}
\UICm{\neg{}P(t)}\DP\\
\end{tabular}  
\end{center}
\hspace{0.2cm}\underline{Relational Rules}
\begin{center}
\begin{tabular}{c@{\hspace{1cm}}c}
\AXCm{P(t_1,\ldots{},t_n)}
\AXCm{\neg{}P(s_1,\ldots{},s_n)}\RL{pred}
\BICm{t_1\neq{}s_1\bdots{}t_n\neq{}s_n}\DP &
\AXCm{f(t_1,\ldots{},t_n)\neq{}f(s_1,\ldots{},s_n)}\RL{fun}
\UICm{t_1\neq{}s_1\bdots{}t_n\neq{}s_n}\DP\\\\
\AXCm{R_s(s,t)}\AXCm{\neg{}R_s(u,v)}\RL{sym}
\BICm{t\neq{}u\brsep{}s\neq{}v}\DP &
\AXCm{\neg{}R_r(s,t)}\RLm{\neg_\mathrm{refl}}
\UICm{s\neq{}t}\DP\\\\
\multicolumn{2}{c}{
\AXCm{R_t(s,t)}\AXCm{\neg{}R_t(u,v)}\RL{trans}
\BICm{u\neq{}s,\neg{}R_t(u,s)\brsep{}t\neq{}v,\neg{}R_t(t,v)}\DP}\\\\
\multicolumn{2}{c}{
\AXCm{R_{ts}(s,t)}\AXCm{\neg{}R_{ts}(u,v)}\RL{transsym}
\BICm{v\neq{}s,\neg{}R_{ts}(v,s)\brsep{}t\neq{}u,\neg{}R_{ts}(t,u)}\DP}\\\\
\multicolumn{2}{c}{
\AXCm{s=t}\AXCm{\neg{}R_t(u,v)}\RL{transeq}
\BICm{u\neq{}s,\neg{}R_t(u,s)\brsep\neg{}R_t(u,s),\neg{}R_t(t,v)\brsep{}
t\neq{}v,\neg{}R_t(t,v)}\DP}\\\\
\multicolumn{2}{c}{
\AXCm{s=t}\AXCm{\neg{}R_{ts}(u,v)}\RL{transeqsym}
\BICm{v\neq{}s,\neg{}R_{ts}(v,s)\brsep
\neg{}R_{ts}(v,s),\neg{}R_{ts}(t,u)\brsep{}
t\neq{}u,\neg{}R_{ts}(t,u)}\DP}
\end{tabular}  
\end{center}}}
\caption{Proof Search Rules of \zenon{}}
\label{fig:zenon}
\end{figure}

\subsection{From Axioms to Superdeduction Rules}

As mentioned in Section~\ref{sec:sded}, reasoning modulo a theory in a tableau
method using superdeduction requires to generate new deduction rules from some
axioms of the theory. The axioms which can be considered for superdeduction are
of the form $\forall{}\bar{x}\;(P\Leftrightarrow{}\varphi{})$, where $P$ is
atomic. This specific form of axiom allows us to introduce an orientation of the
axiom from $P$ to $\varphi{}$, and we introduce the notion of proposition
rewrite rule (this notion appears in~\cite{BA07}, from which we borrow the
following definition and notation):

\begin{definition}[Proposition Rewrite Rule]
The notation $R:P\rightarrow{}\varphi{}$ denotes the axiom
$\forall{}\bar{x}\;(P\Leftrightarrow{}\varphi{})$, where $R$ is the name of the
rule, $P$ an atomic proposition, $\varphi{}$ a proposition, and $\bar{x}$ the
free variables of $P$ and $\varphi{}$.
\end{definition}

It should be noted that $P$ may contain first order terms and therefore that
such an axiom is not just a definition. For instance,
$x\in{}\{~y~|~y\in{}a\land{}y\in{}b~\}\rightarrow{}x\in{}a\land{}x\in{}b$ (where
the comprehension set is a first order term) is a proposition rewrite rule.

Let us now describe how the computation of superdeduction rules for \zenon{} is
performed from a given proposition rewrite rule.

\begin{definition}[Computation of Superdeduction Rules]
Let ${\cal S}$ be a set of rules composed by the subset of the proof search
rules of \zenon{} formed of the closure rules, the analytic rules, as well as
the $\gamma_{\forall{}M}$ and $\gamma_{\neg\exists{}M}$ rules. Given a
proposition rewrite rule $R:P\rightarrow{}\varphi{}$, two superdeduction rules
(a positive one $R$ and a negative one $\neg{}R$) are generated in the following
way:

\begin{enumerate}
\item To get the positive rule $R$, initialize the procedure with the formula
$\varphi{}$. Next, apply the rules of ${\cal S}$ until there is no open leaf
anymore on which they can be applied. Then, collect the premises and the
conclusion, and replace $\varphi$ by $P$ to obtain the positive rule $R$.

\item To get the negative rule $\neg{}R$, initialize the procedure with the
formula $\neg{}\varphi{}$. Next, apply the rules of ${\cal S}$ until there is no
open leaf anymore on which they can be applied. Then, collect the premises and
the conclusion, and replace $\neg{}\varphi$ by $\neg{}P$ to obtain the negative
rule $\neg{}R$.
\end{enumerate}

If the rule $R$ (resp. $\neg{}R$) involves metavariables, an instantiation rule
$R_\mathrm{inst}$ (resp. $\neg{}R_\mathrm{inst}$) is added, where one or several
metavariables can be instantiated.
\end{definition}

Integrating these new deduction rules to the proof search rules of \zenon{} is
sound as they are generated from a subset of rules of \zenon{}, while cut-free
completeness cannot be preserved in general (i.e. for any theory).  In practice,
soundness can be ensured by the ability of \zenon{} of producing proofs for some
proof assistants, such as \coq{} and \isabelle{}, which can be used as proof
checkers.

Let us illustrate the computation of superdeduction rules from a proposition
rewrite rule with the example of the set inclusion.

\begin{example}[Set Inclusion]
From the definition of the set inclusion, we introduce the proposition rewrite
rule $\mathrm{Inc}:a\subseteq{}b\rightarrow{}
\forall{}x\;(x\in{}a\Rightarrow{}x\in{}b)$, and the corresponding superdeduction
rules $\mathrm{Inc}$ and $\neg{}\mathrm{Inc}$ are generated as follows:

\begin{center}
\begin{tabular}{cp{0.5cm}c}
\AXCm{\forall{}x\;(x\in{}a\Rightarrow{}x\in{}b)}\RLm{\gamma_{\forall{}M}}
\UICm{X\in{}a\Rightarrow{}X\in{}b}\RLm{\beta_\Rightarrow}
\UICm{X\not\in{}a\brsep X\in{}b}\DP &&
\AXCm{\neg\forall{}x\;(x\in{}a\Rightarrow{}x\in{}b)}\RLm{\delta_{\neg\forall}}
\UICm{\neg(\epsilon_x\in{}a\Rightarrow{}\epsilon_x\in{}b)}
\RLm{\alpha_{\neg\Rightarrow}}
\UICm{\epsilon_x\in{}a,\epsilon_x\not\in{}b}\DP
\end{tabular}
\end{center}

where $\epsilon_x=\epsilon(x).\neg(x\in{}a\Rightarrow{}x\in{}b)$.

The resulting superdeduction rules are then the following:

\begin{center}
\begin{tabular}{cp{0.5cm}cp{0.5cm}c}
\AXCm{a\subseteq{}b}\RLm{\mathrm{Inc}}
\UICm{X\not\in{}a\brsep X\in{}b}\DP &&
\AXCm{a\subseteq{}b}\RLm{\mathrm{Inc}_\mathrm{inst}}
\UICm{t\not\in{}a\brsep t\in{}b}\DP &&
\AXCm{a\not\subseteq{}b}\RLm{\neg{}\mathrm{Inc}}
\UICm{\epsilon_x\in{}a,\epsilon_x\not\in{}b}\DP
\end{tabular}
\end{center}
\end{example}


\section{An Implementation for the \bmth{} Set Theory}
\label{sec:bset}

In this section, we describe our first extension of \zenon{} with superdeduction
in the case of the \bmth{} set theory, which is the underlying theory of the
\bmth{} formal method, and where superdeduction rules are hard-coded from the
initial axiomatic theory. We also propose an example of \bmth{} proof rule
coming from the database maintained by \siemens{}, and that can be verified by
this tool producing a quite comprehensible proof.

\subsection{Superdeduction Rules for the \bmth{} Set Theory}

This extension of \zenon{} for the \bmth{} set theory is actually motivated by
an experiment which is managed by \siemens{} regarding the verification of
\bmth{} proof rules~\cite{JA11,JA12}. The \bmth{} method~\cite{B-Book}, or
\bmth{} for short, allows engineers to develop software with high guarantees of
confidence; more precisely it allows them to build correct by design
software. \bmth{} is a formal method based on set theory and theorem proving,
and which relies on a refinement-based development process. The \atelierb{}
environment~\cite{Atelier-B} is a platform that supports \bmth{} and offers,
among other tools, both automated and interactive provers. In practice, to
ensure the global correctness of formalized applications, the user must
discharge proof obligations. These proof obligations may be proved
automatically, but otherwise, they have to be handled manually either by using
the interactive prover, or by adding new proof rules that the automated prover
can exploit. These new proof rules can be seen as axioms and must be verified by
other means, otherwise the global correctness may be endangered.

In~\cite{JA11}, we develop an approach based on the use of \zenon{} to verify
\bmth{} proof rules. The method used in this approach consists in first
normalizing the formulas to be proved, in order to obtain first order formulas
containing only the membership set operator, and then calling \zenon{} on these
new formulas. This experiment gives satisfactory results in the sense that it
can prove a significant part of the rules coming from the database maintained by
\siemens{} (we can deal with about 1,400~rules, 1,100~of which can be proved
automatically, over a total of 5,300~rules). However, this approach is not
complete (after the normalization, \zenon{} proves the formulas without any
axiom of set theory, while some instantiations may require to be normalized),
and suffers from efficiency issues (due to the preliminary normalization). To
deal with these problems, the idea developed in~\cite{JA12} is to integrate the
axioms and constructs of the \bmth{} set theory into the \zenon{} proof search
method by means of superdeduction rules. This integration is concretely achieved
thanks to the extension mechanism offered by \zenon{}, which allows us to extend
its core of deductive rules to match specific requirements. This new tool has
emphasized significant speed-ups both in terms of proof time and proof size
compared to the previous approach (see~\cite{JA12} for more details).

The \bmth{} method is based on a typed set theory. There are two rule systems:
one for demonstrating that a formula is well-typed, and one for demonstrating
that a formula is a logical consequence of a set of axioms. The main aim of the
type system is to avoid inconsistent formulas, such as Russell's paradox for
example. The \bmth{} proof system is based on a sequent calculus with equality.
Six axiom schemes define the basic operators and the extensionality which, in
turn, defines the equality of two sets. In addition, the other operators
($\cup{}$, $\cap{}$, etc.) are defined using the previous basic ones. To
generate the superdeduction rules corresponding to the axioms and constructs, we
use the algorithm described in Section~\ref{sec:sded}, and we must therefore
identify the several proposition rewrite rules. Regarding the axioms, they are
all of the form $P_i\Leftrightarrow{}Q_i$, and the associated proposition
rewrite rules are therefore $R_i:P_i\rightarrow{}Q_i$, where each axiom is
oriented from left to right. For instance, let us consider the equality of two
sets, which is defined by the following axiom:

$$a=b\Leftrightarrow{}\forall{}x\;(x\in{}a\Leftrightarrow{}x\in{}b)$$

From this axiom, we can compute two superdeduction rules as follows (a third
rule dealing with instantiation is also implicitly computed since one of the
generated rules involves metavariables):

\begin{center}
\begin{tabular}{cp{0.5cm}c}
\AXCm{a=b}\RLm{=}
\UICm{X\not\in{}a,X\not\in{}b\brsep{}X\in{}a,X\in{}b}\DP &&

\begin{tabular}{c}
\AXCm{a\neq{}b}\RLm{\neq}
\UICm{\epsilon_x\not\in{}a,\epsilon_x\in{}b\brsep{}\epsilon_x\in{}a,
\epsilon_x\not\in{}b}\DP\smallskip{}\\
{\scriptsize with
$\epsilon_x=\epsilon(x).\lnot{}(x\in{}a\Leftrightarrow{}x\in{}b)$}
\end{tabular}
\end{tabular}
\end{center}

Concerning the constructs, they are expressed by definitions of the form
$E_i\triangleq{}F_i$, where $E_i$ and $F_i$ are expressions, and the
corresponding proposition rewrite rules are
$R_i:x\in{}E_i\rightarrow{}x\in{}F_i$. Let us illustrate the computation of
superdeduction rules for constructs with the example of domain restriction,
which is defined in the following way:

$$a\lhd{}b\triangleq{}\mathrm{id}(a);b$$

where:

\begin{center}
\begin{tabular}{l}
$a;b\triangleq{}\{~(x,z)~|~\exists{}y\;((x,y)\in{}a\land{}(y,z)\in{}b~\}$\\
$\mathrm{id}(a)\triangleq{}\{~(x,y)~|~(x,y)\in{}a\times{}a\land{}x=y~\}$
\end{tabular}
\end{center}

The corresponding superdeduction rules are computed as follows:

\begin{center}
\begin{tabular}{cp{0.5cm}c}
\AXC{$(x,y)\in{}a\lhd{}b$}\RL{$\lhd$}
\UIC{$(x,y)\in{}b$, $x\in{}a$}\DP &&

\AXC{$(x,y)\not\in{}a\lhd{}b$}\RL{$\lnot{}\lhd$}
\UIC{$(x,y)\not\in{}b\brsep{}x\not\in{}a$}\DP
\end{tabular}
\end{center}

For further details regarding the computation of superdeduction rules for the
\bmth{} set theory, as well as the corresponding implementation using \zenon{},
the reader can refer to~\cite{JA12}.

\subsection{Verification of a \bmth{} Proof Rule}

To assess our extension of \zenon{} for the \bmth{} set theory using
superdeduction and to show that it can produce proofs comprehensible enough to
recover the intuition of these proofs, we propose to consider the example of a
\bmth{} proof rule coming from the database maintained by \siemens{}. The rule
being considered is the rule named ``SimplifyRelDorXY.27'' (this rule is
actually part of the \atelierb{} set of rules), and whose proof is small enough
to be understood easily and described in the space restrictions for this paper.
The definition of this rule is the following:

$$\emptyset{}\lhd{}a=\emptyset{}$$

When applied to this rule, our extension of \zenon{} produces the proof of
Figure~\ref{fig:bset} (\zenon{} proposes several proof formats, and the proof
presented in this figure uses the format with the highest level of abstraction).
The statement of the rule, i.e. the command starting with ``fof'', is provided
using the TPTP syntax~\cite{TPTP}, and the proof (if found) is displayed after
this statement. The proof consists of several numbered steps, where each of them
is a set of formulas together with a proof rule which has been applied to the
considered proof step. Formulas of a proof step are signed formulas, and
formulas starting with ``-.'' are negative formulas. A proof rule appears at the
end of a proof step and after the string ``\#\#\#'', and also provides, after
the string ``-->'', the other proof steps to which it is connected (these other
proof steps represent the result of the application of this proof rule to the
set of formulas of the considered proof step). For example, in this proof,
Step~1 is connected to Steps~2 and~3. This connection between proof steps
provides the proof with a tree-like structure, where proof steps with axiomatic
rules, i.e. starting with ``Axiom'', are leaves, while the other proof steps are
nodes. Among these other proof steps, there are in particular superdeduction
rules, which start with ``Extension''. In this proof, the \bmth{} constructs are
prefixed by ``b\_'', and ``b\_empty'', ``b\_BIG'', ``b\_in'', ``b\_eq'', and
``b\_drest'' respectively represent the empty set $\emptyset{}$, the set
$\mathrm{BIG}$ (which is an infinite set, mostly only used to build natural
numbers in the foundational theory), the membership operator ``$\in$'', the
(extensional) equality ``$=$'', and the domain restriction construct
``$\lhd{}$''.

\setlength{\codeindent}{0.4cm}

\begin{figure}[t]
\cprotect\framebox
{\begin{minipage}{\textwidth-2\fboxsep-2\fboxrule}
\bigskip{}
\hspace{\codeindent}\begin{minipage}{0cm}
\begin{code}
fof(simplifyRelDorXY_27, conjecture,
  b_eq (b_drest (b_empty, a), b_empty)).
(* PROOF-FOUND *)
   1. H0: (-. (b_eq (b_drest (b_empty) (a)) (b_empty)))
      ### [Extension/b/b_not_eq H0 H1 H2 H3 H4 H5 H6] --> 2 3
   2. H2: (b_in T_7 (b_empty))
      ### [Extension/b/b_in_empty H2 H8 H9 H7] --> 4
   4. H9: (-. (b_in T_7 (b_BIG)))
      H8: (b_in T_7 (b_BIG))
      ### [Axiom H8 H9]
   3. H3: (b_in T_7 (b_drest (b_empty) (a)))
      ### [Extension/b/b_in_drest H3 H10 H11 H12 H7 H6 H13] --> 5
   5. H12: (b_in T_14 (b_empty))
      ### [Extension/b/b_in_empty H12 H15 H16 H14] --> 6
   6. H16: (-. (b_in T_14 (b_BIG)))
      H15: (b_in T_14 (b_BIG))
      ### [Axiom H15 H16]
\end{code}
\end{minipage}
\bigskip{}
\end{minipage}}
\caption{Proof of Rule ``SimplifyRelDorXY.27'' of \atelierb{}}
\label{fig:bset}
\end{figure}

As can be observed, this proof expressed in this format can be easily understood
not only thanks to the tableau method which follows a natural way to find the
proof in this case, but also thanks to superdeduction rules which allow us to
shorten the proof removing formal details useless for the comprehension of the
proof. To justify this claim, let us describe the formal proof sketch which can
be extracted from this formal proof, and which is appropriate to provide the
intuition of the proof. This formal proof sketch is built as follows:

\begin{enumerate}
\item The proof starts from the sequent
``$\vdash{}\emptyset{}\lhd{}a=\emptyset{}$'', which corresponds to Hypothesis~H0
in Step~1 in the formal proof (where the initial formula has been negated since
the tableau method works by refutation). The proof rule applied to this sequent
is a superdeduction rule which deals with the equality, and which corresponds to
the superdeduction rule named ``b\_not\_eq'' in the formal proof, i.e. the
negation of the equality still because the initial formula has been negated. In
sequent calculus, this superdeduction rule is the following:

$$\AXCm{\Gamma{},x\in{}a\vdash{}x\in{}b,\Delta{}}
\AXCm{\Gamma{},x\in{}b\vdash{}x\in{}a,\Delta{}}
\RLm{\mathrm{{=}R}\mbox{, }x\not\in{}\Gamma{},\Delta{}}
\BICm{\Gamma{}\vdash{}a=b,\Delta{}}\DP$$

\item Applying the superdeduction rule for equality, we obtain two cases to
prove (as shown by the rule above). The formal proof focuses on the right-hand
side of the rule at first, and we therefore have to prove the sequent
``$x\in{}\emptyset{}\vdash{}x\in{}\emptyset{}\lhd{}a$'', which corresponds to
Step~2. As can be seen, in the set of formulas of each proof step, the formal
proof only displays the formulas which are useful to complete the proof. For
instance, in Step~2, the formula ``$x\in{}\emptyset{}\lhd{}a$'' is not displayed
(even though it is present in the set of formulas), because it is not used in
the following of the proof. The formal proof therefore focuses on the hypothesis
``$x\in{}\emptyset{}$'' and applies the superdeduction rule named
``b\_in\_empty'' and corresponding to the empty set. In \bmth{}, the empty set
is defined as follows: $\emptyset{}\triangleq{}\mathrm{BIG}-\mathrm{BIG}$. In
sequent calculus, the corresponding superdeduction rule is then computed as
follows:

$$\AXCm{\Gamma{},x\in{}\mathrm{BIG},x\not\in{}\mathrm{BIG}\vdash{}\Delta{}}
\RLm{\emptyset{}\mathrm{L}}
\UICm{\Gamma{},x\in{}\emptyset{}\vdash{}\Delta{}}\DP$$

Once this rule has been applied, we obtain the sequent
``$x\in{}\mathrm{BIG},x\not\in{}\mathrm{BIG}\vdash{}x\in{}\emptyset{}\lhd{}a$'',
which is proved by reductio ad absurdum and corresponds to Step~4 in the formal
proof.

\item The second case following the application of the superdeduction rule for
equality corresponds to the sequent
``$x\in{}\emptyset{}\lhd{}a\vdash{}x\in{}\emptyset{}$'', which appears to be
Step~3 in the formal proof. In this step, the formal proof focuses on the
hypothesis ``$x\in{}\emptyset{}\lhd{}a$'', and applies the superdeduction rule
named ``b\_in\_drest'' and corresponding to the domain restriction. This
superdeduction rule is the following in sequent calculus:

$$\AXCm{\Gamma{},x=(y,z),(y,z)\in{}b,y\in{}a\vdash{}\Delta{}}
\RLm{\lhd{}\mathrm{L}\mbox{, }y,z\not\in{}\Gamma{},\Delta{}}
\UICm{\Gamma{},x\in{}a\lhd{}b\vdash{}\Delta{}}\DP$$

Once this rule has been applied, we obtain the sequent
``$x=(y,z),(y,z)\in{}a,y\in{}\emptyset{}\vdash{}x\in{}\emptyset{}$'', where the
formal proof can again focus on the hypothesis $y\in{}\emptyset{}$ in Step~5 and
close the proof as previously in Step~6.
\end{enumerate}

From this formal proof sketch, it is now quite easy to produce an informal and
short proof (as it would have been done in a textbook) as follows:

\begin{itemize}
\item To show $\emptyset{}\lhd{}a=\emptyset{}$, we have to consider two cases:
given $x\in{}\emptyset{}$, we must show $x\in{}\emptyset{}\lhd{}a$, and given
$x\in{}\emptyset{}\lhd{}a$, we must show $x\in{}\emptyset{}$;

\begin{enumerate}
\item If $x\in{}\emptyset{}$ then $x\in{}\mathrm{BIG}$ and
$x\not\in{}\mathrm{BIG}$, which is therefore absurd;

\item If $x\in{}\emptyset{}\lhd{}a$ then $x=(y,z)$, $(y,z)\in{}a$, and
$y\in{}\emptyset{}$, which is absurd as previously.
\end{enumerate}
\end{itemize}


\section{A Generic Implementation for First Order Theories}
\label{sec:szen}

In this section, we present our second extension of \zenon{} with
superdeduction, which is able to deal with any first order theory. In this
extension, the theory is analyzed to determine the axioms which can be turned
into superdeduction rules, and these superdeduction rules are automatically
computed on the fly to enrich the deductive kernel of \zenon{}. We also describe
the proofs of two examples coming from the TPTP library and produced by this
tool, and which are quite comprehensible as well.

\subsection{From Theories to Superdeduction Systems}

This extension of \zenon{} is actually a generalization of the previous one
dedicated to the \bmth{} set theory, where superdeduction rules are henceforth
automatically computed on the fly. In the previous extension, superdeduction
rules are hard-coded since the \bmth{} set theory is a higher order theory due
to one of the axioms of the theory (the comprehension scheme), and we have to
deal with this axiom specifically in the implementation of \zenon{}. Even though
some techniques exist to handle higher order theories as first order theories
(like the theory of classes, for example), a hard-coding of these theories may
be preferred as these techniques unfortunately tend to increase the entropy of
the proof search. In addition, in the previous extension, some of the
superdeduction rules must be manually generated as they must be shrewdly tuned
(ordering the several branches of the rules, for instance) to make the tool
efficient. The new extension of \zenon{} dealing with any first order theory has
been developed as a tool called \szen{}~\cite{Super-Zenon}, where each theory is
analyzed to determine the axioms which are candidates to be turned into
superdeduction rules. As said in Section~\ref{sec:auto}, axioms of the form
$\forall{}\bar{x}\;(P\Leftrightarrow{}\varphi{})$, where $P$ is atomic, can be
transformed, but we can actually deal with more axioms. Here is the exhaustive
list of axioms that can be handled, as well as the corresponding superdeduction
rules that can be generated (in the following, $P$ and $P'$ are atomic, and
$\varphi{}$ is an arbitrary formula):

\begin{itemize}
\item Axiom of the form $\forall{}\bar{x}\;(P\Leftrightarrow{}\varphi{})$: we
consider the proposition rewrite rule $R:P\rightarrow{}\varphi{}$, and the two
superdeduction rules $R$ and $\neg{}R$ are generated;

\item Axiom of the form $\forall{}\bar{x}\;(P\Rightarrow{}P')$: we
consider the proposition rewrite rules $R:P\rightarrow{}P'$ and
$R':\neg{}P'\rightarrow{}\neg{}P$, and only the superdeduction rules $R$ and
$R'$ are generated;

\item Axiom of the form $\forall{}\bar{x}\;(P\Rightarrow{}\varphi{})$: we
consider the proposition rewrite rule $R:P\rightarrow{}\varphi{}$, and only the
superdeduction rule $R$ is generated;

\item Axiom of the form $\forall{}\bar{x}\;(\varphi{}\Rightarrow{}P)$: we
consider the proposition rewrite rule $R:\neg{}P\rightarrow{}\neg{}\varphi{}$,
and only the superdeduction rule $R$ is generated;

\item Axiom of the form $\forall{}\bar{x}\;P$: we consider the proposition
rewrite rule $R:\neg{}P\rightarrow{}\bot{}$, and only the superdeduction rule
$R$ is generated.
\end{itemize}

The axioms of the theory, which are not of these forms, are left as regular
axioms. An axiom, which is of one of these forms, is also left as a regular
axiom if the conclusion of one of the generated superdeduction rules (i.e. the
top formula of one of these rules) unifies with the conclusion of an already
computed superdeduction rule (in this case, the theory is actually
non-deterministic and we try to minimize this source of non-determinism by
dividing these incriminated axioms among the sets of superdeduction rules and
regular axioms). An axiom, which is of one of these forms, is still left as a
regular axiom if $P$ is an equality (as we do not want to interfere with the
specific management of equality by the kernel of \zenon{}). Finally, it should
be noted that for axioms of the form $\forall{}\bar{x}\;(P\Rightarrow{}P')$, we
also consider the proposition rewrite rule which corresponds to the converse of
the initial formula; this actually allows us to keep cut-free completeness in
this particular case.

\subsection{Proof of a Logic Puzzle}

As the \szen{} tool is able to deal with any first order theory, it can be used
in many contexts, and in particular, it can be applied to all the first order
problems of the TPTP library~\cite{TPTP} (about 6,600~problems), which is a
library of test problems for automated theorem proving systems. To assess the
effectiveness of this tool and to show that it can also produce comprehensible
proofs, let us consider an example of the puzzle category of TPTP and called
``Crime in Beautiful Washington'' (Puzzle \#132), which is a problem in the same
vein as the ``Who Killed Aunt Agatha?'' well-known puzzle. This kind of problems
is quite appropriate for educational purposes when teaching artificial
intelligence and logic for example. The problem being considered consists of the
following axioms:

$$\begin{array}{ll}
\forall{}x\;(\mathrm{capital}(x)\Rightarrow{}\mathrm{city}(x)) &
(\mathit{capital\_city\_type})\\
\mathrm{capital}(\mathrm{washington}) & (\mathit{washington\_type})\\
\mathrm{country}(\mathrm{usa}) & (\mathit{usa\_type})\\
\forall{}x\;(\mathrm{country}(x)\Rightarrow{}
\mathrm{capital}(\mathrm{capital\_city}(x))) &
(\mathit{country\_capital\_type})\\
\forall{}x\;(\mathrm{city}(x)\Rightarrow{}\mathrm{has\_crime}(x)) &
(\mathit{crime\_axiom})\\
\mathrm{capital\_city}(\mathrm{usa})=\mathrm{washington} &
(\mathit{usa\_capital\_axiom})\\
\forall{}x\;(\mathrm{country}(x)\Rightarrow{}
\mathrm{beautiful}(\mathrm{capital\_city}(x))) &
(\mathit{beautiful\_capital\_axiom})
\end{array}$$

As can be observed, all the axioms can be turned into superdeduction rules,
except the axiom $(\mathit{usa\_capital\_axiom})$ since it is an equality, and
the axiom $(\mathit{beautiful\_capital\_axiom})$ since one of the generated
superdeduction rules for this axiom overlap with one of the superdeduction rules
computed previously for the axiom $(\mathit{country\_capital\_type})$.

The conjecture to be proved is expressed as follows:

$$\mathrm{beautiful}(\mathrm{washington})\land{}
\mathrm{has\_crime}(\mathrm{washington})$$

When applied to this specification, \szen{} produces the proof of
Figure~\ref{fig:puz} for the previous conjecture (we still use the proof format
with the highest level of abstraction). As in the example of verification of a
\bmth{} proof rule in Section~\ref{sec:bset}, it is possible to build quite
directly the following informal proof sketch from this formal proof:

\begin{itemize}
\item To show $\mathrm{beautiful}(\mathrm{washington})\land{}
\mathrm{has\_crime}(\mathrm{washington})$, we have to consider the two cases
$\mathrm{beautiful}(\mathrm{washington})$ and
$\mathrm{has\_crime}(\mathrm{washington})$;

\begin{enumerate}
\item To show $\mathrm{beautiful}(\mathrm{washington})$, we apply the axiom
$(\mathit{beautiful\_capital\_axiom})$ instantiated with $\mathrm{usa}$, and we
have to consider two cases: we must show $\mathrm{country}(\mathrm{usa})$, and
given $\mathrm{beautiful}(\mathrm{capital\_city}(usa)$, we must show
$\mathrm{beautiful}(\mathrm{washington})$;

\begin{enumerate}
\item To show $\mathrm{country}(\mathrm{usa})$, we use the axiom
$(\mathit{usa\_type})$;

\item Given $\mathrm{beautiful}(\mathrm{capital\_city}(usa)$, to show
$\mathrm{beautiful}(\mathrm{washington})$, it is enough to show
$\mathrm{capital\_city}(\mathrm{usa})=\mathrm{washington}$ using the axiom
$(\mathit{usa\_capital\_axiom})$.
\end{enumerate}

\item To show $\mathrm{has\_crime}(\mathrm{washington})$, we apply the axiom
$(\mathit{crime\_axiom})$, and we must show
$\mathrm{city}(\mathrm{washington})$;

\begin{itemize}
\item To show $\mathrm{city}(\mathrm{washington})$, we apply the axiom
$(\mathit{capital\_city\_type})$, and we must show
$\mathrm{capital}(\mathrm{washington})$;

\item To show $\mathrm{capital}(\mathrm{washington})$, we use the axiom
$(\mathit{washington\_type})$.
\end{itemize}
\end{enumerate}
\end{itemize}

\begin{figure}[t]
\cprotect\framebox
{\begin{minipage}{\textwidth-2\fboxsep-2\fboxrule}
\bigskip{}
\hspace{\codeindent}\begin{minipage}{0cm}
\begin{code}
fof(washington_conjecture, conjecture,
  (beautiful (washington) & has_crime (washington))).
(* PROOF-FOUND *)
   1. H0: (-. ((beautiful (washington)) /\ (has_crime (washington))))
      H1: ((capital_city (usa)) = (washington))
      H2: (All X, ((country X) => (beautiful (capital_city X))))
      ### [NotAnd H0] --> 2 3
   2. H3: (-. (beautiful (washington)))
      ### [All H2] --> 4
   4. H4: ((country (usa)) => (beautiful (capital_city (usa))))
      ### [Imply H4] --> 5 6
   5. H5: (-. (country (usa)))
      ### [Extension/szen/usa_type H5]
   6. H6: (beautiful (capital_city (usa)))
      ### [P-NotP H6 H3] --> 7
   7. H7: ((capital_city (usa)) != (washington))
      ### [Axiom H1 H7]
   3. H8: (-. (has_crime (washington)))
      ### [Extension/szen/crime_axiom H8 H9 H10] --> 8
   8. H9: (-. (city (washington)))
      ### [Extension/szen/capital_city_type H9 H11 H10] --> 9
   9. H11: (-. (capital (washington)))
      ### [Extension/szen/washington_type H11]
\end{code}
\end{minipage}
\bigskip{}
\end{minipage}}
\caption{Proof of Puzzle \#132 of TPTP}
\label{fig:puz}
\end{figure}

\subsection{Proof of a Geometry Problem}

As a second example of proof, we consider a problem coming from the geometry
category of the TPTP library. This problem (Problem \#170+3) states that if two
distinct points are incident with a line, then this line is equivalent with the
connecting line of these points. The interest of such an example is actually
twofold. First, the corresponding proof is larger (but remains reasonably large
to be presented in this paper) than for the previous considered examples, which
tends to show that our approach is effective even when the proofs require more
than 20~steps. Second, the subject of geometry is quite fundamental in the high
school curriculum in the sense that it is generally the only mathematical topic
where proofs are explicitly mentioned and where formal reasoning is actually
considered. The axioms considered in this example are the axioms of constructive
geometry, but \szen{} uses classical logic and constructive geometry plus
classical logic is equivalent to textbook theories. The proof of this example
uses the following axioms of this theory (we use the names given in the TPTP
files):

$$\begin{array}{ll}
\forall{}x,y\;(\mathrm{distinct\_points}(x,y)\Rightarrow{}\neg{}
\mathrm{apart\_point\_and\_line}(x,\mathrm{line\_connecting}(x,y))) &
(\mathit{ci1})\\
\forall{}x,y\;(\mathrm{distinct\_points}(x,y)\Rightarrow{}\neg{}
\mathrm{apart\_point\_and\_line}(y,\mathrm{line\_connecting}(x,y))) &
(\mathit{ci2})\\
\forall{}x,y,u,v\;(\mathrm{distinct\_points}(x,y)\land{}
\mathrm{distinct\_lines}(u,v)\Rightarrow{}\\
\phantom{\forall{}x,y,u,v\;(}\mathrm{apart\_point\_and\_line}(x,u)\lor{}
\mathrm{apart\_point\_and\_line}(x,v)\lor{}\\
\phantom{\forall{}x,y,u,v\;(}\mathrm{apart\_point\_and\_line}(y,u)\lor{}
\mathrm{apart\_point\_and\_line}(y,v)) & (\mathit{cu1})\\
\forall{}x,y\;(\mathrm{equal\_lines}(x,y)\Leftrightarrow{}
\neg{}\mathrm{distinct\_lines}(x,y)) & (\mathit{ax2})\\
\forall{}x,y\;(\mathrm{incident\_point\_and\_line}(x,y)\Leftrightarrow{}
\neg{}\mathrm{apart\_point\_and\_line}(x,y)) & (\mathit{a4})
\end{array}$$

where $\mathrm{distinct\_points}(x,y)$ (resp. $\mathrm{distinct\_lines}(x,y)$)
means that $x$ and $y$ are two distinct points (resp. lines),
$\mathrm{incident\_point\_and\_line}(x,y)$
(resp. $\mathrm{apart\_point\_and\_line}(x,y)$) means that the point $x$ is
(resp. is not) incident with the line $y$, $\mathrm{equal\_lines}(x,y)$ means
that $x$ and $y$ denote the same line, and $\mathrm{line\_connecting}(x,y)$
denotes the line connecting the points $x$ and $y$.

Among these axioms, the axioms $(\mathit{ci2})$, $(\mathit{ax2})$, and
$(\mathit{a4})$ are turned into superdeduction rules. The axiom $(\mathit{ci1})$
is left as an axiom because one of its superdeduction rules overlap with one of
the superdeduction rules computed previously for the axiom $(\mathit{ci2})$. The
axiom $(\mathit{cu1})$ is also left as an axiom because it has not the right
form to be turned into superdeduction rules. It should be noted that for the
axiom $(\mathit{ax2})$, a superdeduction rule corresponding to the converse of
this axiom is also generated since both sides of the implication are atomic.

The conjecture given previously is formally expressed as follows:

$$\begin{array}{l}
\forall{}x,y,z\;(\mathrm{distinct\_points}(x,y)\land{}
\mathrm{incident\_point\_and\_line}(x,z)\land{}\\
\phantom{\forall{}x,y,z\;(}\mathrm{incident\_point\_and\_line}(y,z)\Rightarrow{}
\mathrm{equal\_lines}(z,\mathrm{line\_connecting}(x,y)))
\end{array}$$

When applied to this problem, \szen{} is able to produce the proof of
Figures~\ref{fig:geo1} and~\ref{fig:geo2} (we use the same proof format than for
the previous examples), where the preliminary Skolemization steps are compressed
(Steps~2 and~3 are left implicit). From this proof, it is possible to build the
following informal proof sketch as previously:

\begin{figure}[p]
\cprotect\framebox
{\begin{minipage}{\textwidth-2\fboxsep-2\fboxrule}
\bigskip{}
\hspace{\codeindent}\begin{minipage}{0cm}
\begin{code}
fof(geometry_conjecture, conjecture,
  (! [X, Y, Z] : ((distinct_points (X, Y) &
    incident_point_and_line (X, Z) &
    incident_point_and_line (Y, Z)) =>
    equal_lines (Z, line_connecting(X, Y))))).
(* PROOF-FOUND *)
   1. H0: (-. (All X, (All Y, (All Z, (((distinct_points X Y) /\
            ((incident_point_and_line X Z) /\
            (incident_point_and_line Y Z))) =>
            (equal_lines Z (line_connecting X Y)))))))
      H1: (All X, (All Y, ((distinct_points X Y) =>
            (-. (apart_point_and_line X (line_connecting X Y))))))
      H2: (All X, (All Y, (All U, (All V, (((distinct_points X Y) /\
            (distinct_lines U V)) => ((apart_point_and_line X U) \/
            ((apart_point_and_line X V) \/
            ((apart_point_and_line Y U) \/
            (apart_point_and_line Y V)))))))))
      ### [NotAllEx H0] --> [...] 4
   4. H7: (incident_point_and_line T_4 T_8)
      H9: (-. (equal_lines T_8 (line_connecting T_4 T_6)))
      H10: (distinct_points T_4 T_6)
      H11: (incident_point_and_line T_6 T_8)
      ### [Extension/szen/a4 H7 H12 H4 H8] --> 5
   5. H12: (-. (apart_point_and_line T_4 T_8))
      ### [Extension/szen/a4 H11 H13 H6 H8] --> 6
   6. H13: (-. (apart_point_and_line T_6 T_8))
      ### [Extension/szen/not_ax2 H9 H14 H8 H15] --> 7
   7. H14: (distinct_lines T_8 (line_connecting T_4 T_6))
      ### [All H2] --> 8
   8. H16: (All Y, (All U, (All V, (((distinct_points T_4 Y) /\
             (distinct_lines U V)) => ((apart_point_and_line T_4 U) \/
             ((apart_point_and_line T_4 V) \/
             ((apart_point_and_line Y U) \/
             (apart_point_and_line Y V))))))))
      ### [All H16] --> 9
   9. H17: (All U, (All V, (((distinct_points T_4 T_6) /\
             (distinct_lines U V)) => ((apart_point_and_line T_4 U) \/
             ((apart_point_and_line T_4 V) \/
             ((apart_point_and_line T_6 U) \/
             (apart_point_and_line T_6 V)))))))
      ### [All H17] --> 10
\end{code}
\end{minipage}
\bigskip{}
\end{minipage}}
\caption{Proof of Geometry Problem \#170+3 of TPTP (Part~1)}
\label{fig:geo1}
\end{figure}

\begin{figure}[p]
\cprotect\framebox
{\begin{minipage}{\textwidth-2\fboxsep-2\fboxrule}
\bigskip{}
\hspace{\codeindent}\begin{minipage}{0cm}
\begin{code}
  10. H18: (All V, (((distinct_points T_4 T_6) /\
             (distinct_lines T_8 V)) =>
             ((apart_point_and_line T_4 T_8) \/
             ((apart_point_and_line T_4 V) \/
             ((apart_point_and_line T_6 T_8) \/
             (apart_point_and_line T_6 V))))))
      ### [All H18] --> 11
  11. H19: (((distinct_points T_4 T_6) /\
             (distinct_lines T_8 (line_connecting T_4 T_6))) =>
             ((apart_point_and_line T_4 T_8) \/
             ((apart_point_and_line T_4 (line_connecting T_4 T_6)) \/
             ((apart_point_and_line T_6 T_8) \/
             (apart_point_and_line T_6 (line_connecting T_4 T_6))))))
      ### [DisjTree H19] --> 12 13 14 15 16 17
  12. H20: (-. (distinct_points T_4 T_6))
      ### [Axiom H10 H20]
  13. H21: (-. (distinct_lines T_8 (line_connecting T_4 T_6)))
      ### [Axiom H14 H21]
  14. H22: (apart_point_and_line T_4 T_8)
      ### [Axiom H22 H12]
  15. H23: (apart_point_and_line T_4 (line_connecting T_4 T_6))
      ### [All H1] --> 18
  18. H24: (All Y, ((distinct_points T_4 Y) =>
             (-. (apart_point_and_line T_4 (line_connecting T_4 Y)))))
      ### [All H24] --> 19
  19. H25: ((distinct_points T_4 T_6) =>
             (-. (apart_point_and_line T_4 (line_connecting T_4 T_6))))
      ### [Imply H25] --> 20 21
  20. H20: (-. (distinct_points T_4 T_6))
      ### [Axiom H10 H20]
  21. H26: (-. (apart_point_and_line T_4 (line_connecting T_4 T_6)))
      ### [Axiom H23 H26]
  16. H27: (apart_point_and_line T_6 T_8)
      ### [Axiom H27 H13]
  17. H28: (apart_point_and_line T_6 (line_connecting T_4 T_6))
      ### [Extension/szen/ci2ctrp H28 H20 H4 H6] --> 22
  22. H20: (-. (distinct_points T_4 T_6))
      ### [Axiom H10 H20]
\end{code}
\end{minipage}
\bigskip{}
\end{minipage}}
\caption{Proof of Geometry Problem \#170+3 of TPTP (Part~2)}
\label{fig:geo2}
\end{figure}

\begin{itemize}
\item Given the points $x$, $y$, and the line $z$ s.t.
$\mathrm{distinct\_points}(x,y)$, $\mathrm{incident\_point\_and\_line}(x,z)$,
and $\mathrm{incident\_point\_and\_line}(y,z)$, we have to show
$\mathrm{equal\_lines}(z,\mathrm{line\_connecting}(x,y))$;

\item From the hypotheses $\mathrm{incident\_point\_and\_line}(x,z)$ and
$\mathrm{incident\_point\_and\_line}(y,z)$, we have
$\neg{}\mathrm{apart\_point\_and\_line}(x,z)$ and
$\neg{}\mathrm{apart\_point\_and\_line}(y,z)$ using the axiom $(\mathit{a4})$;

\item Using the axiom $(\mathit{ax2})$, the goal 
$\mathrm{equal\_lines}(z,\mathrm{line\_connecting}(x,y))$ to be proved is
equivalent to
$\neg{}\mathrm{distinct\_lines}(z,\mathrm{line\_connecting}(x,y))$;

\item Using the axiom $(\mathit{cu1})$ with $x$, $y$, $z$, and
$\mathrm{line\_connecting}(x,y)$, we have to show the previous goal in the
following cases:

\begin{enumerate}
\item Given $\neg{}\mathrm{distinct\_points}(x,y)$, we have also
$\mathrm{distinct\_points}(x,y)$ in hypothesis, which is therefore absurd;

\item Given $\neg{}\mathrm{distinct\_lines}(z,\mathrm{line\_connecting}(x,y))$,
it is exactly the goal to be proved, which is then proved directly by
hypothesis;

\item Given $\mathrm{apart\_point\_and\_line}(x,z)$, we have also
$\neg{}\mathrm{apart\_point\_and\_line}(x,z)$ in hypothesis, which is therefore
absurd;

\item Given
$\mathrm{apart\_point\_and\_line}(x,\mathrm{line\_connecting}(x,y))$,
we use the axiom $(\mathit{ci1})$ with $x$, $y$, and
$\mathrm{distinct\_points}(x,y)$, to have
$\neg{}\mathrm{apart\_point\_and\_line}(x,\mathrm{line\_connecting}(x,y))$,
which is therefore absurd;

\item Given $\mathrm{apart\_point\_and\_line}(y,z)$, we have also
$\neg{}\mathrm{apart\_point\_and\_line}(y,z)$ in hypothesis, which is therefore
absurd;

\item Given $\mathrm{apart\_point\_and\_line}(y,\mathrm{line\_connecting}(x,y))$
used with the converse of the axiom $(\mathit{ci2})$ with $x$ and $y$, we have
$\neg{}\mathrm{distinct\_points}(x,y)$, which is therefore absurd as we have
also $\mathrm{distinct\_points}(x,y)$ in hypothesis.
\end{enumerate}
\end{itemize}


\section{Conclusion}

In this paper, we have proposed an automated deduction method which allows us to
produce proofs close to the human intuition and practice. This method is based
on tableaux and uses the principles of superdeduction, among which the theory is
used to enrich the deduction system with new deduction rules, called
superdeduction rules. We have presented two implementations of this method,
which consist of extensions of the \zenon{} automated theorem prover. The first
implementation is a version dedicated to the \bmth{} set theory, where the
superdeduction rules are hard-coded from the initial axiomatic theory. The
second implementation is a generic version able to deal with any first order
theory, where the theory is analyzed to determine the axioms which can be turned
into superdeduction rules, and where these superdeduction rules are
automatically computed on the fly to enrich the deductive kernel of
\zenon{}. For information, these two implementations are available as free
software at~\cite{Super-Zenon}. We have also provided some examples of problems,
which can be handled by these tools and which come from different theories, such
as the \bmth{} set theory or theories of the TPTP library (in the puzzle and
geometry categories, in particular). In these examples, we have shown that both
tools are able to produce formal proofs comprehensible enough to recover the
intuition of these proofs, and that the user can therefore easily extract
informal proof sketches from these proofs.

As future work, it would be interesting to improve the readability of the
produced proofs in order to get more natural proofs and in particular, it might
be desirable to turn proofs into pure direct proofs (searching for a proof of
the initial formula), rather than refutational proofs (searching to invalidate
the negation of the initial formula). In a way, this corresponds to get back to
Gentzen's initial purely proof theoretical motivation when trying to find proofs
in the cut-free version of sequent calculus, and in particular, this is opposed
to Hintikka and Beth's semantic view of tableaux, which consists of a procedure
systematically trying to find a counter example for a given formula (i.e. a
model in which its negation is true). Such an improvement evokes some similar
initiatives in other automated theorem provers, such as
\muscadet{}~\cite{Muscadet}, which is based on natural deduction and which uses
methods resembling those used by humans.

To increase the readability of the proofs generated by our extensions, it would
also be interesting to export these proofs to other kinds of languages, which
appear more appropriate regarding readability. In particular, we could export
the proofs to declarative proof languages, such as \isar{}~\cite{MW99} for
\isabelle{}, which tends to bridge the semantic gap between internal notions of
proof given by state-of-the-art interactive theorem proving systems and an
appropriate level of abstraction for user-level work. This translation should be
automatic, and to be more effective, it should also be probably combined with an
interactive layer over the automated deduction tool (see below) in order to
produce intelligible proofs, i.e. proofs where a certain number of cuts can be
manually introduced. We could even go further and automatically produce proofs
in natural languages using, for example, the ideas of~\cite{YC97}, where \coq{}
formal proofs are translated in a pseudo natural language.

Finally, in this paper, our extensions only produce proofs automatically without
any interaction with the user. In an educational setting, a system able to
present sample proofs is already a valuable bonus, but the students must also be
involved in the process of building proofs. To do so, the idea is to implement
an interactive layer over our extensions in the spirit of~\cite{DV06}, which
will aim to offer the user the possibility to guide the proof search. This
interactive layer would be a benefit for both the user and the automated
deduction tool. For the user, this layer could make the interface with the proof
engine, which could propose a set of applicable rules or next-step hints. For
the automated deduction tool, this layer could be used to find proofs with the
help of the user, who could propose to focus on some branches of the proof
search, which would allow the tool to find a proof, while the strategy of the
tool would have focused on inappropriate branches resulting in an endless proof
search.

\bibliographystyle{plain}
\bibliography{biblio}

\begin{thebibliography}{10}

\bibitem{B-Book}
Jean-Raymond Abrial.
\newblock {\em The \bbook{}, Assigning Programs to Meanings}.
\newblock Cambridge University Press, Cambridge (UK), 1996.
\newblock ISBN 0521496195.

\bibitem{EB55}
Evert~Willem Beth.
\newblock {Semantic Entailment and Formal Derivability}.
\newblock {\em Medelingen von de Koninklijke Nederlandse Akademie van
  Wetenschappen, Afdeling Letterkunde}, 18(13):309--342, 1955.

\bibitem{Zenon}
Richard Bonichon, David Delahaye, and Damien Doligez.
\newblock {\zenon{}: An Extensible Automated Theorem Prover Producing Checkable
  Proofs}.
\newblock In {\em Logic for Programming Artificial Intelligence and Reasoning
  (LPAR)}, volume 4790 of {\em LNCS/LNAI}, pages 151--165, Yerevan (Armenia),
  October 2007. Springer.

\bibitem{BA07}
Paul Brauner, Clément Houtmann, and Claude Kirchner.
\newblock {Principles of Superdeduction}.
\newblock In {\em Logic in Computer Science (LICS)}, pages 41--50, Wroc\l{}aw
  (Poland), July 2007. IEEE Computer Society Press.

\bibitem{RB86}
Randal~E. Bryant.
\newblock {Graph-Based Algorithms for Boolean Function Manipulation}.
\newblock {\em IEEE Transactions on Computers}, 35(8):677--691, August 1986.

\bibitem{GB11}
Guillaume Burel.
\newblock {Efficiently Simulating Higher-Order Arithmetic by a First-Order
  Theory Modulo}.
\newblock {\em Logical Methods in Computer Science (LMCS)}, 7(1):1--31, March
  2011.

\bibitem{Atelier-B}
\clearsy{}.
\newblock {\em \atelierb{}~4.0}, February 2009.
\newblock \url{http://www.atelierb.eu/}.

\bibitem{YC97}
Yann Coscoy.
\newblock {A Natural Language Explanation for Formal Proofs}.
\newblock In {\em Logical Aspects of Computational Linguistics (LACL)}, volume
  1328 of {\em LNCS}, pages 149--167, Nancy (France), September 1996. Springer.

\bibitem{DP60}
Martin Davis and Hillary Putnam.
\newblock {A Computing Procedure for Quantification Theory}.
\newblock {\em Journal of the Association for Computing Machinery (JACM)},
  7(3):201--215, 1960.

\bibitem{DA03}
Gilles Dowek, Thérèse Hardin, and Claude Kirchner.
\newblock {Theorem Proving Modulo}.
\newblock {\em Journal of Automated Reasoning (JAR)}, 31(1):33--72, September
  2003.

\bibitem{GG35}
Gerhard Gentzen.
\newblock {Untersuchungen über das Logische Schlie\ss{}en}.
\newblock {\em Mathematische Zeitschrift}, 39(2/3):176--210, 405--431, 1935.

\bibitem{KH55}
K.~Jaakko~J. Hintikka.
\newblock {Form and Content in Quantification Theory}.
\newblock {\em Acta Philosophica Fennica}, 8:7--55, 1955.

\bibitem{JA11}
Mélanie Jacquel, Karim Berkani, David Delahaye, and Catherine Dubois.
\newblock {Verifying \bmth{} Proof Rules using Deep Embedding and Automated
  Theorem Proving}.
\newblock In {\em Software Engineering and Formal Methods (SEFM)}, volume 7041
  of {\em LNCS}, pages 253--268, Montevideo (Uruguay), November 2011. Springer.

\bibitem{JA12}
Mélanie Jacquel, Karim Berkani, David Delahaye, and Catherine Dubois.
\newblock {Tableaux Modulo Theories using Superdeduction: An Application to the
  Verification of \bmth{} Proof Rules with the \zenon{} Automated Theorem
  Prover}.
\newblock In {\em International Joint Conference on Automated Reasoning
  (IJCAR)}, volume 7364 of {\em LNCS}, pages 332--338, Manchester (UK), June
  2012. Springer.

\bibitem{Super-Zenon}
Mélanie Jacquel and David Delahaye.
\newblock {\em \szen{}, version~0.0.1}.
\newblock \siemensr{} and \cnam{}, May 2012.
\newblock \url{http://cedric.cnam.fr/~delahaye/super-zenon/}.

\bibitem{Muscadet}
Dominique Pastre.
\newblock {\em \muscadet{}~4.1}.
\newblock Université Paris Descartes (Paris~5), April 2011.
\newblock \\\url{http://www.math-info.univ-paris5.fr/~pastre/muscadet/}.

\bibitem{DP65}
Dag Prawitz.
\newblock {Natural Deduction. A Proof-Theoretical Study}.
\newblock {\em Stockholm Studies in Philosophy}, 3, 1965.

\bibitem{JR65}
John~Alan Robinson.
\newblock {A Machine-Oriented Logic Based on the Resolution Principle}.
\newblock {\em Journal of the Association for Computing Machinery (JACM)},
  12(1):23--41, January 1965.

\bibitem{TPTP}
Geoff Sutcliffe.
\newblock {The TPTP Problem Library and Associated Infrastructure: The FOF and
  CNF Parts, v3.5.0}.
\newblock {\em Journal of Automated Reasoning (JAR)}, 43(4):337--362, December
  2009.

\bibitem{Coq}
{The \coq{} Development Team}.
\newblock {\em \coq{}, version~8.4}.
\newblock \inria{}, August 2012.
\newblock \\\url{http://coq.inria.fr/}.

\bibitem{DV06}
Daniel~J. Velleman.
\newblock {\em How to Prove It: A Structured Approach}.
\newblock Cambridge University Press, New York (NY, USA), April 2006.
\newblock ISBN 9780521675994.

\bibitem{MW99}
Markus Wenzel.
\newblock {\isar{} -- A Generic Interpretative Approach to Readable Formal
  Proof Documents}.
\newblock In {\em Theorem Proving in Higher Order Logics (TPHOLs)}, volume 1690
  of {\em LNCS}, pages 167--184, Nice (France), September 1999. Springer.

\end{thebibliography}

\end{document}